# Two-color flat-top solitonic pulses in $\chi^{(2)}$ optical microresonators via second harmonic generation

Valery E. Lobanov[1], Nikita M. Kondratiev[1], Artem E. Shitikov[1,2],

Igor A. Bilenko[1,2]

[1]*Russian Quantum Center, Skolkovo 143025, Russia*

[2]*Faculty of Physics, Lomonosov Moscow State University, Moscow 119991, Russia*

**Abstract**

We studied numerically the generation of the coherent frequency combs at second harmonic generation in $\chi^{(2)}$ microresonators via conventional frequency scan method. It was demonstrated for the first time that under particular conditions it is possible to generate two-color flat-top solitonic pulses, platicons, using pump amplitude modulation or controllable mode interaction approach, if the signs of the group velocity coefficients at pump frequency and its second harmonic are opposite but absolute values of these coefficients are rather close. It was revealed that platicons may be observed on both sides of the linear microresonator resonance (at positive, as well as negative pump frequency detunings). For the efficient platicon excitation, one needs simultaneous accurate matching of both microresonator free spectral ranges and resonant eigenfrequencies. Platicon generation processes were simulated numerically, excitation conditions and platicon generation domains were found for different generation methods, and the properties of generated platicons were studied for the different combinations of the medium parameters.

During the last decade generation of Kerr frequency combs was demonstrated in high-Q microresonators of different geometries, bulk, and on-chip, made of different materials (crystalline fluorides, diamond, quartz, silicon, silicon nitride, etc.) [1-4]. Coherent microresonator-based frequency combs or dissipative Kerr solitons [5, 6] were found to be an efficient tool for various important applications, such as metrology [7, 8], spectroscopy [9, 10], astrophysics [11, 12], high volume telecommunications [13]. Recently, it was shown that the generation of the optical frequency combs is also possible in non-centrosymmetric materials possessing quadratic nonlinearity, such as $LiNbO_3$ or $LiTaO_3$ [14-20]. Such materials are widely available and well-studied for frequency conversion and development of different photonic devices such as microresonator-based optical modulators and optical parametric oscillators [2, 21-30]. Generation of optical frequency combs due to the quadratic nonlinearity may be realized at reduced pump powers because of the high level of nonlinear response and in spectral diapasons inaccessible for conventional Kerr frequency combs. Also, it was shown that quadratic nonlinearity may support different types of the dissipative solitons (bright, dark and quasi-solitons) in optical microresonators [31, 32]. In previous studies the main attention was paid to the existence domains and properties of the localized states but not to their generation. However, recent studies of the dark solitons in Kerr microresonators at normal group velocity dispersion (GVD) showed that the existence of the stationary solutions of the particular equations does not guarantee the possibility of their generation in real experiments and one should often elaborate original experimental approaches [33, 34]. In our work using the model elaborated in [20, 31] for doubly resonant microresonator SHG systems we simulated the process of the frequency comb generation via frequency scan that is the conventional method of the dissipative Kerr solitons generation in experiments [5]. Bright soliton generation was observed when the group velocity dispersions (GVD) for the pump frequency and its second harmonic had the same signs. In the case of different dispersion signs, at some parameter range the excitation of the localized states was not observed despite the prediction of their existence in [31, 32]. However, those works only predict that the solitonic solutions exist, but do not show how they can be actually generated under realistic experimental conditions. In our work we demonstrated numerically for the first time that applying the methods developed for the platicon generation in Kerr microresonators [33-36], we can obtain the generation of the two-color flat-top solitonic pulses, platicons, in $\chi^{(2)}$ optical microresonators. Platicon excitation was observed for both combinations of the GVD coefficients with the opposite signs. Unpredictably, platicon generation may occur at both positive and negative pump frequency detunings from linear microresonator resonance. Excitation conditions and platicon generation domains were found for different generation methods, and the properties of generated platicons were studied for the different combinations of the medium parameters.

For numerical analysis we used the system of the two coupled equations for the fundamental wave (FW) and second harmonic (SH) fields [20, 31] which, in normalized form, may be written as

$$\begin{cases} \dfrac{\partial u}{\partial \tau} = i\dfrac{1}{2}b_{21}\dfrac{\partial^2 u}{\partial \varphi^2} + ivu^* - (1+i\alpha_1)u + f, \\ \dfrac{\partial v}{\partial \tau} + d\dfrac{\partial v}{\partial \varphi} = i\dfrac{1}{2}b_{22}\dfrac{\partial^2 v}{\partial \varphi^2} + iu^2 - (\dfrac{\kappa_2}{\kappa_1} + i\alpha_2)v, \end{cases} \quad (1)$$

where $u$ and $v$ are the normalized slowly varying envelopes of the FW and SH fields, respectively. Here $\tau = \kappa_1 t/2$ denotes the normalized time, $\kappa_{1,2} = \omega_{01,02}/Q_{1,2}$ denotes the FW (SH) cavity decay rate, $\omega_{01}$ is the microresonator eigenfrequency, closest to the pump frequency $\omega_p$, $\omega_{02}$ is the microresonator eigenfrequency, nearest to the doubled pump frequency, $Q_{1,2}$ is the total quality factor at the fundamental frequency/second harmonic, $\varphi \in [-\pi; \pi]$ is an azimuthal angle in a coordinate system rotating with the angular frequency equal to the microresonator free spectral range (FSR) $D_{11}$ at the fundamental frequency, $d = 2(D_{12} - D_{11})/\kappa_1$ is the normalized difference between FSRs at the fundamental and second harmonic frequencies (temporal walk-off term), $\alpha_1 = 2(\omega_{01} - \omega_p)/\kappa_1$ is the normalized pump frequency detuning, $\alpha_2 = 2(\omega_{02} - 2\omega_p)/\kappa_1 = 2\alpha_1 + 2(\omega_{02} - 2\omega_{01})/\kappa_1 = 2\alpha_1 + \delta$, where $\delta = 2(\omega_{02} - 2\omega_{01})/\kappa_1$ is the normalized offset between the SH resonant frequency $\omega_{02}$ and doubled frequency of the fundamental resonance $\omega_{01}$. $f$ stands for the dimensionless pump amplitude. $b_{21}$ and $b_{22}$ are the GVD coefficients at the fundamental and second harmonic frequencies, respectively. Positive/negative GVD coefficients corresponds here to the anomalous/normal GVD respectively.

We studied numerically nonlinear processes arising upon pump frequency scan across the fundamental frequency resonance ($\omega_p = \omega_p(0) - \Omega t$) with a noise-like input for different combinations of the GVD coefficients $b_{21,22}$. This method is widely used in experiments for the dissipative Kerr solitons generation [5, 6]. Also, it may provide identification of the spectral ranges, where soliton or platicon generation is possible (marked as "solitons" and "platicons" in figures below). Taking into account that an important parameter of Eq. 1 is the difference between the pump frequency and microresonator resonant frequency, and assuming microresonator eigenfrequencies to be constant, in numerical simulations it is convenient to introduce linear in time variation of the pump frequency detuning $\alpha_1$ ($\alpha_1 = \alpha_1(0) + \beta\tau$, $\beta = \frac{4}{\kappa_1^2}\Omega$) and the consequent variation of $\alpha_2 = 2\alpha_1 + \delta$.

Eq. 1 was solved numerically using a standard split-step Fourier routine with 1024 points in the azimuthal direction. We also checked that results do not change with increase of the number of the transverse points. For analysis, we calculated the dependencies $U_{1,2}(\alpha_1)$, where $U_1 = \int_{-\pi}^{\pi} |u|^2 d\varphi$ and $U_2 = \int_{-\pi}^{\pi} |v|^2 d\varphi$ are FW and SH intracavity powers, and studied the field distribution evolution upon frequency scan. As it was shown in [5, 33, 36], a transformation of the dependence $U_1(\alpha_1)$ may indicate the generation of the dissipative localized structures.

As in [31], we set $d = 0$, $f = 10$, $\kappa_2 / \kappa_1 = 1$. In order to guarantee the generation of the steady-state structures and to discriminate them from the transient distributions, normalized frequency scan velocity $\beta$ was chosen rather small ($\beta = 0.001$) and it was checked that dynamics of the considered processes does not change if $\beta$ decreases further. We also considered the case of the ideal matching of the resonant frequencies ($\delta = 0$), thus $\alpha_2 = 2\alpha_1$. Simultaneous matching of FSRs and resonant frequencies can be done by the correct choice of the pumped mode and fine tuning of the microresonator geometry [32, 37]. Phase matching of the microresonator may be realized through periodic poling [2, 24]. It was shown in [32, 37] that resonant conditions for the different polarizations of the interacting waves may correspond to the different combinations of the signs of the GVD coefficients.

First, we studied the case of the equal GVD coefficients (see Fig. 1): $b_{21} = b_{22} = 0.05$ (anomalous GVD) and $b_{21} = b_{22} = -0.05$ (normal GVD). Soliton generation was observed upon frequency scan (see characteristic step at the dependence $U_1(\alpha_1)$) at positive values of the pump detuning for the anomalous GVD and at negative values for the normal GVD. Such soliton existence domains are in good agreement with the predictions arising from the model using Kerr-like cascading nonlinearity which sign depends on the pump detuning sign [32, 38, 39]. Note, that number of generated solitons is not constant and may vary from realization to realization [40].

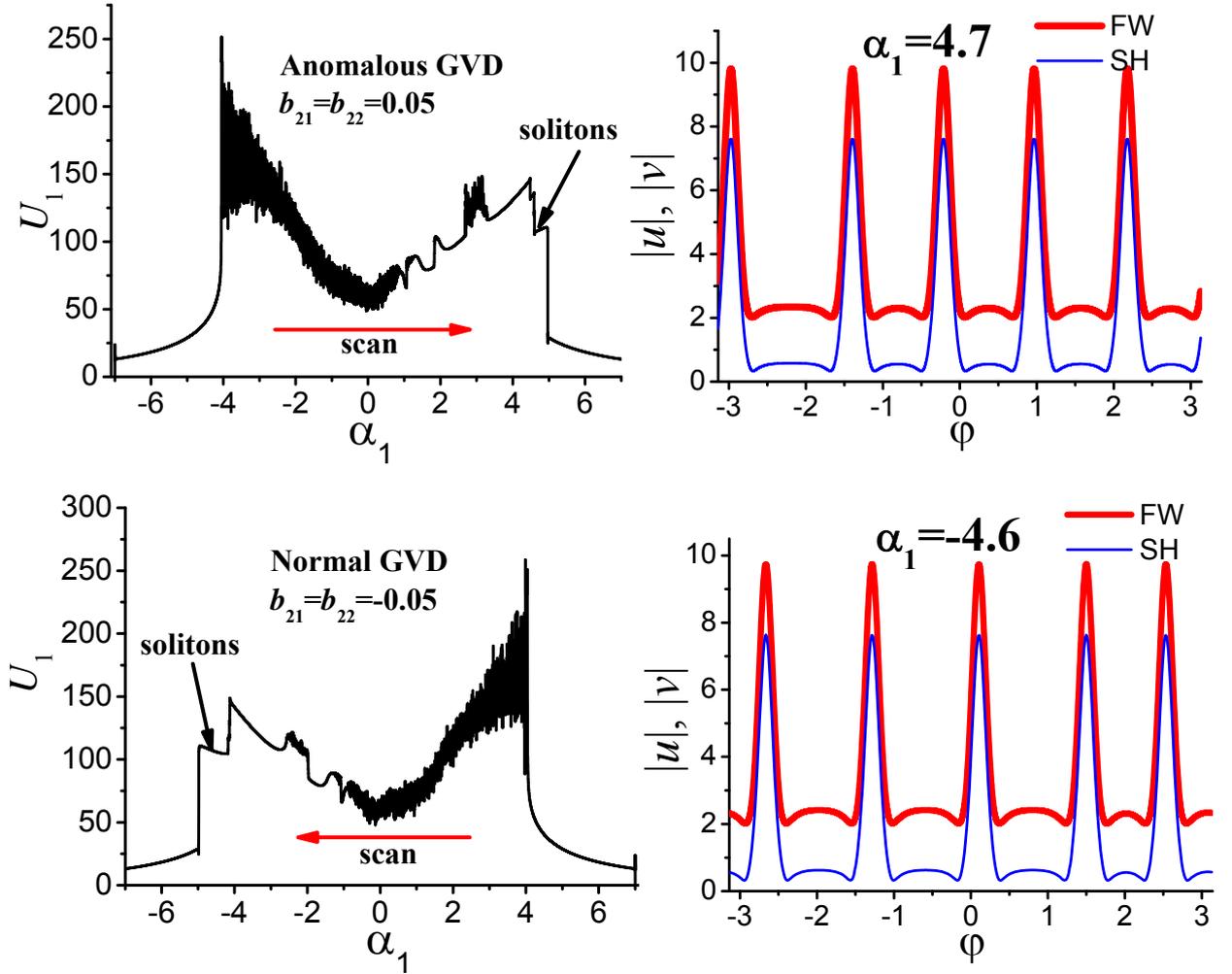

**Fig. 1.** (left column) The dependence $U_1(\alpha_1)$ upon frequency scan at anomalous and normal GVD at both frequencies; (right column) soliton components profiles. All quantities are plotted in dimensionless units.

When GVD signs were different ($b_{21} = 0.05$, $b_{22} = -0.05$ or $b_{21} = -0.05$, $b_{22} = 0.05$) we observed some unstable behavior in the vicinity of the cold resonance point ($\alpha_1 = 0$) but high-intensity branches remained stable upon frequency scan (forward and backward) resulting in the triangle resonant curves at positive and negative detuning values (see Fig. 2). Note, that the same effects were observed for the wide range of pump amplitudes when dispersion coefficients at pump frequency and its second harmonic have opposite signs but rather close absolute values (for $f = 10$, $b_{21} = 0.05$ it takes place if $-0.086 < b_{22} < -0.028$).

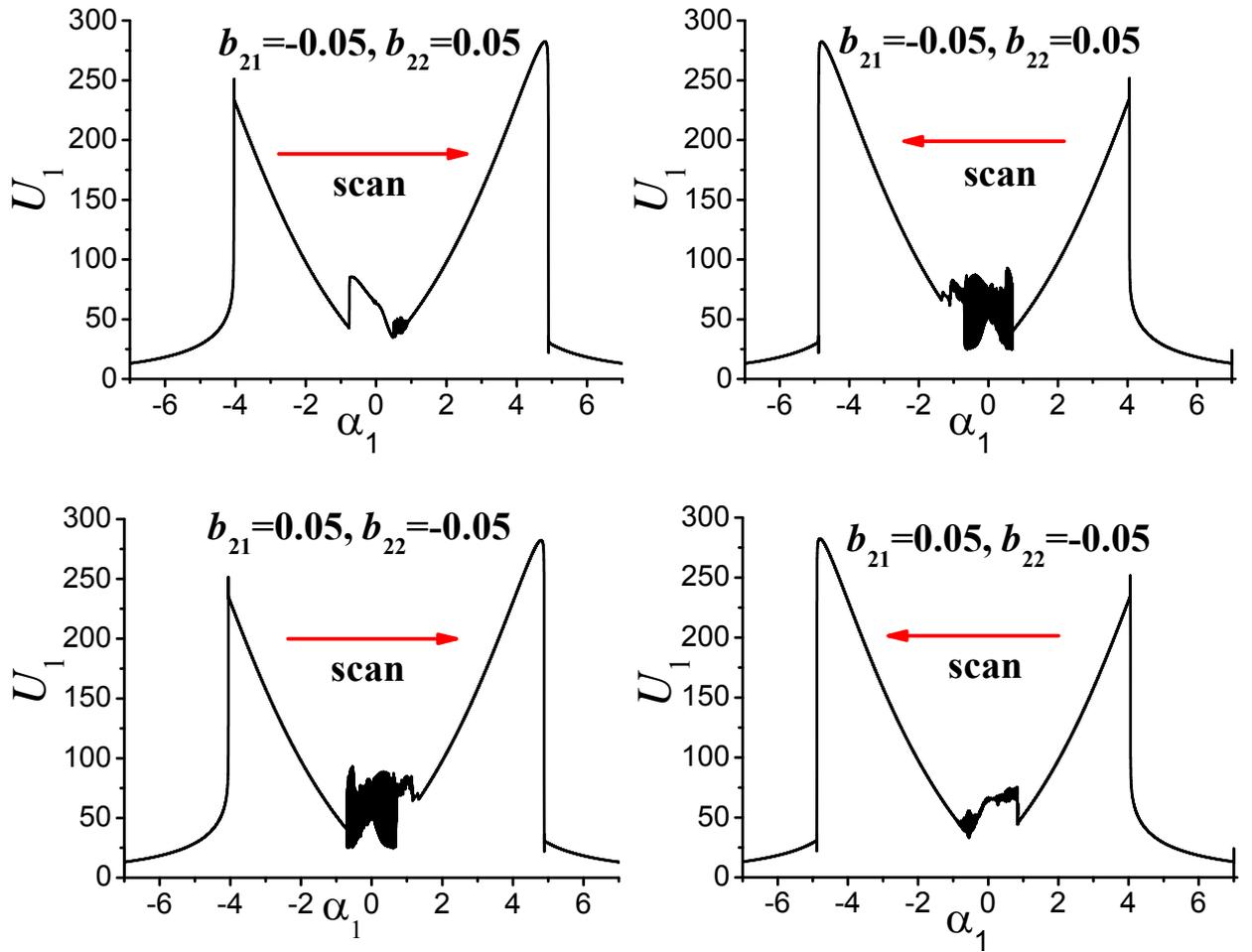

**Fig. 2.** The dependence $U_1(\alpha_1)$ upon forward (left column) and backward (right column) frequency scan at $b_{21} = -0.05$, $b_{22} = 0.05$ (upper panels) and $b_{21} = 0.05$, $b_{22} = -0.05$ (bottom panels). All quantities are plotted in dimensionless units.

However, such stability of the upper branch gives up an opportunity to generate localized structures similar to the dark solitons [41-45] or platicons [33-36, 46] in Kerr microresonators using the methods elaborated for the generation of the coherent Kerr frequency combs in the normal dispersion regime. Interestingly, it was shown that generation of the dark solitons or platicons may be significantly more efficient than the generation of the bright soliton trains in microresonators in terms of the pump-to-comb conversion efficiency that is usually limited to a few percent for bright solitons [47, 48] and may exceed 30% [49] (41% was reported in the recent paper [50]) for the dark pulse mode-locking in the normal dispersion range. This aspect was shown to be particularly promising for coherent optical communications [51,52].

As the first approach, we studied the method based on the pump amplitude modulation at the microresonator FSR [34, 36]. The feasibility of this method for platicon generation in Kerr microresonators was proved experimentally [53]. To take pump modulation into account we replaced the homogeneous pump term $f$ in Eq. 1 by the modulated one $f(1+\varepsilon\cos\varphi)$, where $\varepsilon$ is the modulation depth. One may notice the dramatic transformation of the resonance curve when amplitude modulation is introduced (compare resonance curves for the unmodulated and modulated pump in the top left panel of Fig. 3) indicating platicon generation [34].

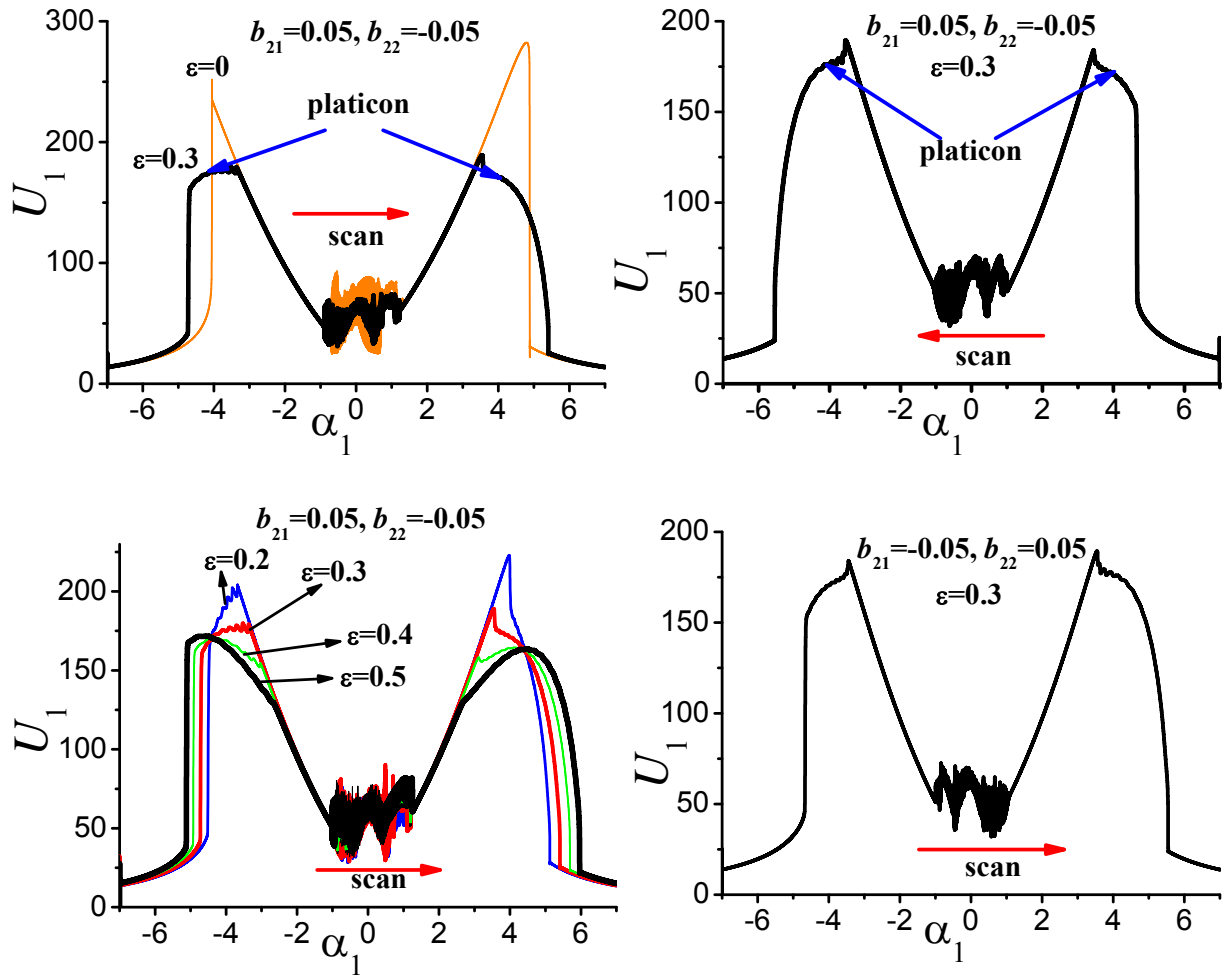

**Fig. 3.** The dependence $U_1(\alpha_1)$ upon forward and backward frequency scan for different values of the modulation depth at $b_{21} = 0.05$, $b_{22} = -0.05$ (top right, top left, bottom left panels) and $b_{21} = -0.05$, $b_{22} = 0.05$ (bottom right panel). All quantities are plotted in dimensionless units.

To verify platicon generation, we also studied field distribution and spectrum evolution upon frequency scan (see Fig. 4). In the upper panels in Fig. 4 one may notice that at some positive and negative detuning values abrupt change of the field distribution takes place, indicating platicon formation. These detuning ranges correspond to the ranges of the resonance curve transformation in Fig. 3. Some unstable behavior without generation of localized structures was observed in the vicinity of the point $\alpha_1 = 0$. In the frequency domain platicon generation manifests itself in a significant broadening of the spectrum (see bottom panels in Fig. 4 where mode numbers $\mu_1$ are defined relative to the pumped mode corresponding to $\mu_1 = 0$ and $\mu_2$ – relative to the mode with the eigenfrequency nearest to the doubled pump frequency).

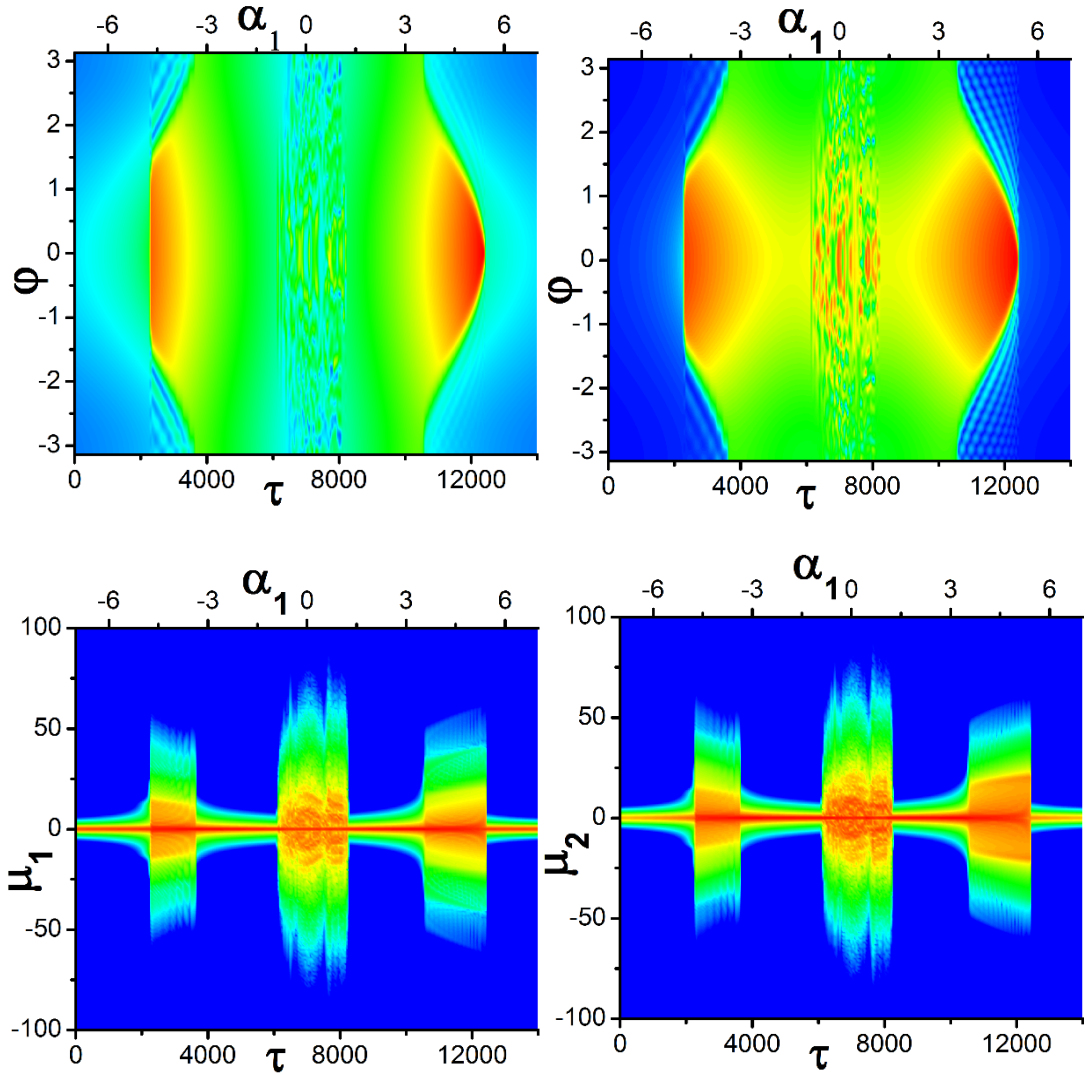

**Fig. 4.** (left column) FW and (right column) SH field distribution (upper line) and spectrum (bottom line) evolution upon forward frequency scan at $b_{21} = 0.05$, $b_{22} = -0.05$, $\varepsilon = 0.3$. All quantities are plotted in dimensionless units.

We also checked that at fixed detuning value generated patterns propagate in a stable fashion over indefinitely large periods of time.

In contrast with Kerr microresonators, in $\chi^{(2)}$ microresonators platicon generation takes place on both sides of the linear microresonator resonance and upon both forward and backward frequency scans (see upper panels in Fig. 3). This fact cannot be predicted using the widely used model of Kerr-like cascading nonlinearity

[32, 38, 39]: in [31] the existence of the dark solitons was predicted only at negative detunings for $b_{21} > 0$ and $b_{22} < 0$, and only at positive detunings for $b_{21} < 0$ and $b_{22} > 0$.

Combining the results of the forward and backward scans we found platicon generation domains for the different values of the modulation depth $\varepsilon$ and pump amplitude $f$ (see Fig. 5). The spectral range of platicon generation becomes wider with the growth of the modulation depth (see the left panel in Fig. 5) and shifts to the larger absolute values of the detuning with the growth of the pump amplitude (see the right panel in Fig. 5).

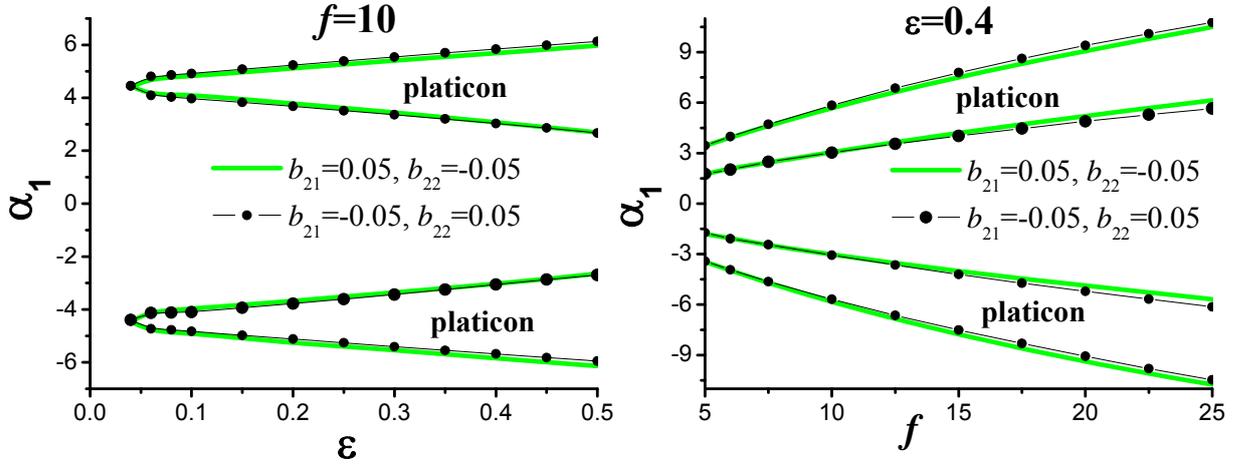

**Fig. 5.** Platicon generation domains for different values of the modulation depth and fixed pump amplitude (left panel) and for different pump amplitudes and fixed modulation depth (right panel) at $b_{21} = 0.05$, $b_{22} = -0.05$ and $b_{21} = -0.05$, $b_{22} = 0.05$. All quantities are plotted in dimensionless units.

Platicons generated at equal absolute values but different signs of pump detunings possess almost equal widths and amplitudes but different types of oscillating tails (see the top panels in Fig. 6).

Interestingly, that practically the same generation dynamics is observed for opposite combinations of the GVD coefficients (compare top left panel in Fig. 3 corresponding to $b_{21} = 0.05$, $b_{22} = -0.05$ and bottom right panel corresponding to $b_{21} = -0.05$, $b_{22} = 0.05$). Platicon generation domains were found to almost coincide for the both combinations of GVD coefficients (compare thick lines and thin lines marked with circles in Fig. 5).

It is shown in Fig. 6 that for both combinations of the GVD coefficients for the same value of the pump detuning $\alpha_1$ platicons have practically the same amplitude and width. Platicons becomes narrower with the growth of the absolute value of the detuning (compare right and left columns in Fig. 6). However, one may notice that the difference is in the oscillating tails: for $b_{21} = 0.05$, $b_{22} = -0.05$ oscillations of the fundamental wave profile are more pronounced at negative detunings, while for $b_{21} = -0.05$, $b_{22} = 0.05$ – at positive detunings (compare top and bottom panels in Fig. 6).

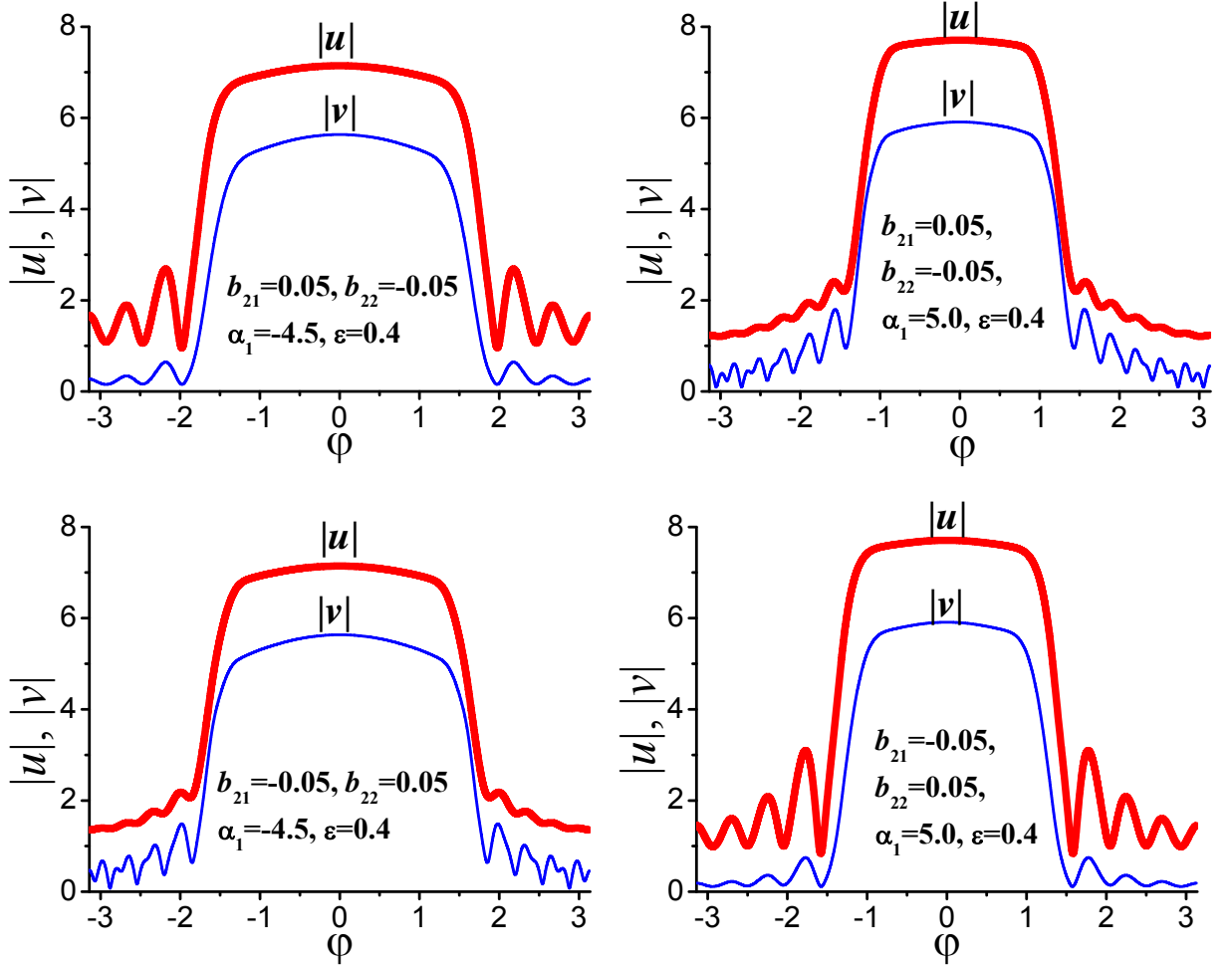

**Fig. 6.** Profiles of the platicon components for different detuning values at $\varepsilon = 0.4$ for $b_{21} = 0.05$, $b_{22} = -0.05$ (top panels) and $b_{21} = -0.05$, $b_{22} = 0.05$ (bottom panels). All quantities are plotted in dimensionless units.

We also found that considered method is applicable for the wide range of the material and pump parameters. Platicon generation was observed for the different pump amplitudes (at least for $f = 5...25$), different absolute values of the GVD coefficients ($|b_{21}| = 0.025...0.25$) and different values of the losses at the second harmonic frequency. Platicon generation range becomes narrower with the growth of the decay rate at the second harmonic $\kappa_2$ and it shifts to the smaller absolute values of the pump detuning $\alpha_1$ (see top left panel in Fig. 7). Also, if $\kappa_2$ increases, platicon profile becomes more smooth and less localized that indicates narrowing of its spectrum (compare platicon profiles in Fig. 7).

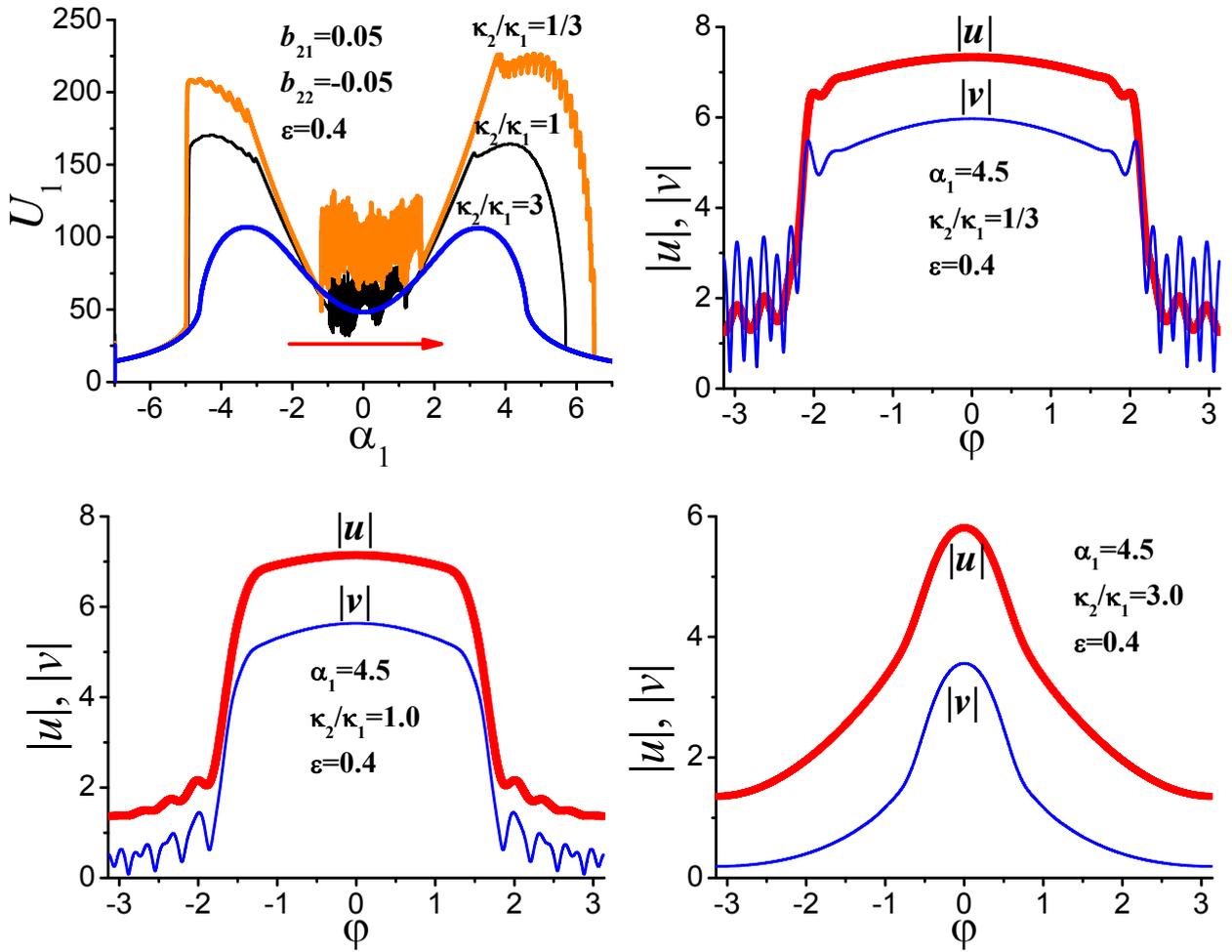

**Fig. 7.** (top left panel) The dependences $U_1(\alpha_1)$ upon frequency scan at $b_{21} = 0.05$, $b_{21} = -0.05$, $\varepsilon = 0.4$ for different values of $\kappa_2 / \kappa_1$; platicon components profiles at $\alpha_1 = 4.5$ for $\kappa_2 / \kappa_1 = 1/3$ (top right panel), $\kappa_2 / \kappa_1 = 1$ (left bottom panel) and $\kappa_2 / \kappa_1 = 3$ (right bottom panel). All quantities are plotted in dimensionless units.

The most simulations were carried out for the same absolute values of FW and SH GVD coefficients ($|b_{21}|=|b_{22}|$). However, it was revealed that platicons can be generated if absolute values of the GVD coefficients are not equal, but rather close. For example, at $f = 10$, $\varepsilon = 0.4$ and $|b_{21}| = 0.05$ platicons were generated if $0.4 < |b_{21}/b_{22}| < 2.5$. Moreover, the range of the valid GVD coefficients depends on the particular combination of their signs and, for example, it is not the same for $b_{21} > 0$, $b_{22} < 0$ and for $b_{21} < 0$, $b_{22} > 0$. Outside this range of parameters, high-intensity branch of the nonlinear resonance may become unstable at some range of detunings, generation range narrows and platicons become indented or disappear. It is well-known that it is possible to tune GVD coefficient by engineering the resonator dispersion via the resonator geometry [54-58], however simultaneous dispersion engineering at FW and SH is not studied yet.

We also checked that this method is quite sensitive to the matching of FSRs at the fundamental and at second harmonic frequencies. Platicon generation may be realized if normalized mismatch $d = 2(D_{12} - D_{11})/\kappa_1$ is less than some critical value depending on the modulation depth. Also, positive mismatch values may lead to the narrowing of the platicon generation range (and even to suppression of generation) at negative detunings $\alpha_1$, negative mismatches – to the same effect at negative detunings. For considered parameters, platicon generation was observed for $d \leq 0.5...1$, thus FSRs difference must be less than $0.25\,\kappa_1...0.5\,\kappa_1$. Note, that even in the presence of the temporal walk-off (if $d \neq 0$) platicon repetition rate is equal to the modulation frequency, but platicon profiles become asymmetric (see Fig. 8).

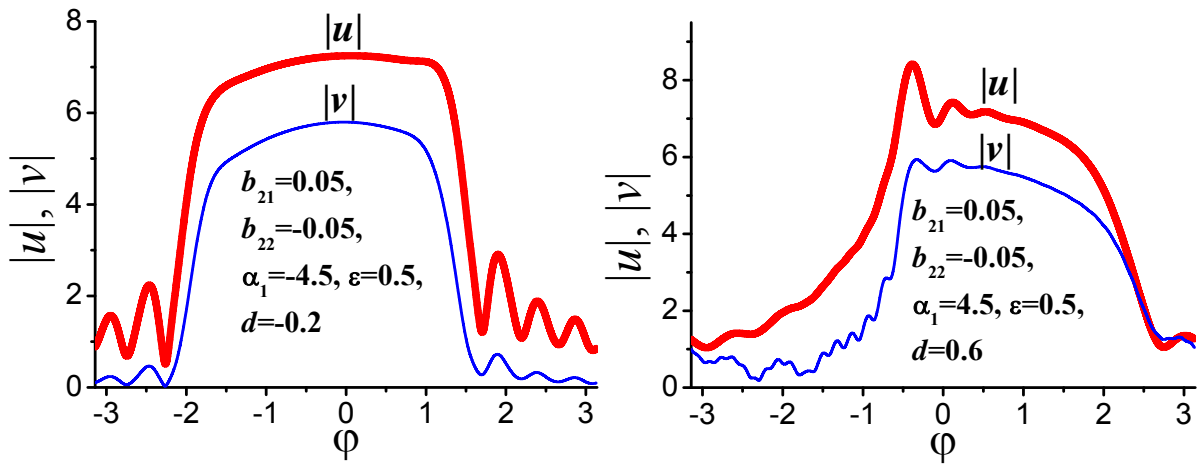

**Fig. 8.** Profiles of the platicon components for different FSRs mismatch values at $\varepsilon = 0.5$ for $b_{21} = 0.05$, $b_{22} = -0.05$. All quantities are plotted in dimensionless units.

The condition of the resonant frequencies matching, $\delta = 0$ or $\omega_{02} = 2\omega_{01}$, should be satisfied rather accurately. If $\delta \neq 0$, platicon amplitudes generated at positive and negative detunings becomes different (see Fig. 9, bottom panels). Generation ranges become asymmetric with respect to the point $\alpha_1 = 0$ (see Fig. 9, top panels). Moreover, depending on the sign of the offset $\delta$, platicon generation may be even suppressed at positive or negative pump detunings $\alpha_1$. At $f = 10$ and

$\varepsilon = 0.4$ platicon generation was observed up to $|\delta| \sim 10$, thus the resonant frequencies offset $|\omega_{02} - 2\omega_{01}|$ should be less than several $\kappa_1$.

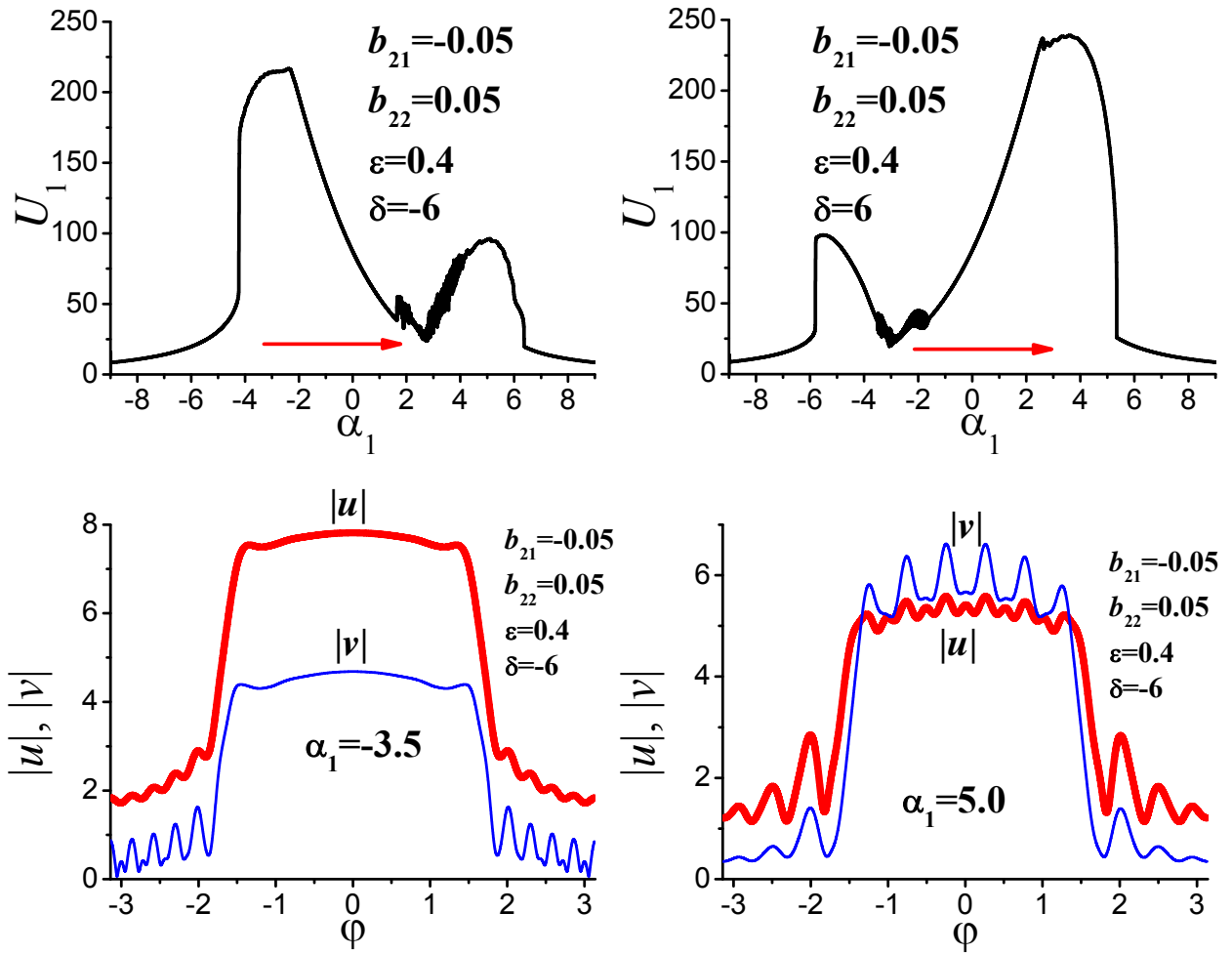

**Fig. 9.** (top) The dependences $U_1(\alpha_1)$ upon frequency scan at $b_{21} = -0.05$, $b_{21} = 0.05$, $\varepsilon = 0.4$ for different values of $\delta$; (bottom) platicon components profiles at $b_{21} = -0.05$, $b_{21} = 0.05$, $\varepsilon = 0.4$, $\delta = -6$ for different values of $\alpha_1$. All quantities are plotted in dimensionless units.

The seconds approach that can be used for platicon generation exploits the coupling between different mode families [43, 59, 60], e.g. in a system of two coupled microresonators [50, 61]. In Kerr microresonators this method can provide pump-to-comb conversion efficiency up to 41% in the in the normal GVD regime [50]. The simplest model for analysis of this effect uses pumped mode shift approximation [33, 35, 46] and describes platicon generation process rather accurately [60].

For numerical analysis, we introduced the frequency shift $\Delta$ of the pumped mode eigenfrequency $\omega_{01}$: $\bar{\omega}_{01} = \omega_{01} - \Delta$. To take it into account, additional phase shift defined by $\Delta$ was applied to the central mode in the frequency domain step of the split-step Fourier routine. Thus, for the pumped mode, pump detuning $\alpha$ should be replaced by the effective detuning $\alpha_{10} = \alpha_1 - (2\Delta / \kappa_1)$.

It was found that if pumped mode shift is introduced platicon, generation may be observed (see characteristic steps at black curves in Fig. 10 and corresponding evolution of the intracavity field distributions and spectra in Fig. 11). Interestingly, at $b_{21} = 0.05$, $b_{22} = -0.05$ platicons may be generated at positive detunings if pumped mode shift is positive and at negative – if it is negative (see Fig. 10).

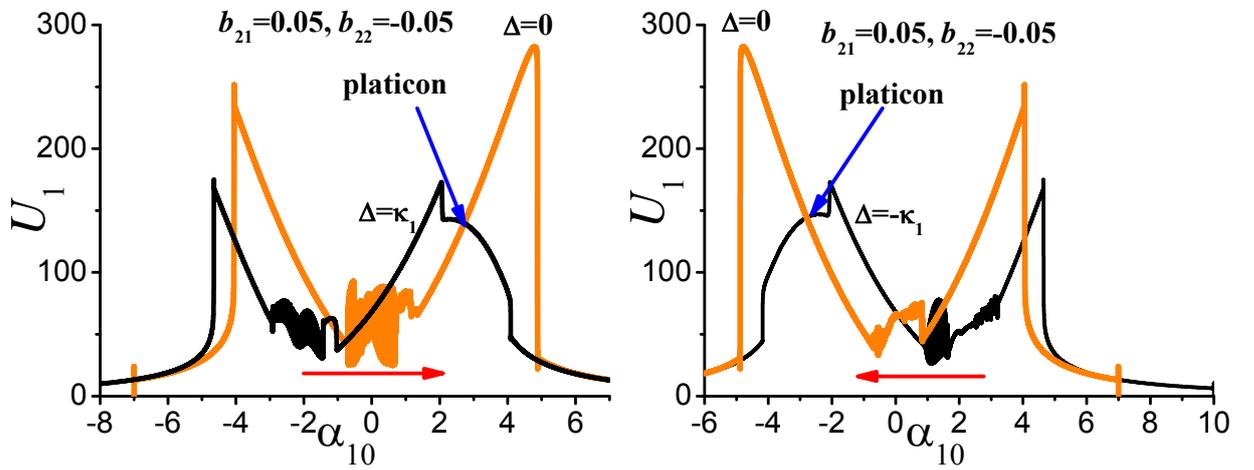

**Fig. 10.** The dependence $U_1(\alpha_{10})$ upon forward and backward frequency scan for different values of pumped mode shift at $b_{21} = 0.05$, $b_{22} = -0.05$. All quantities are plotted in dimensionless units.

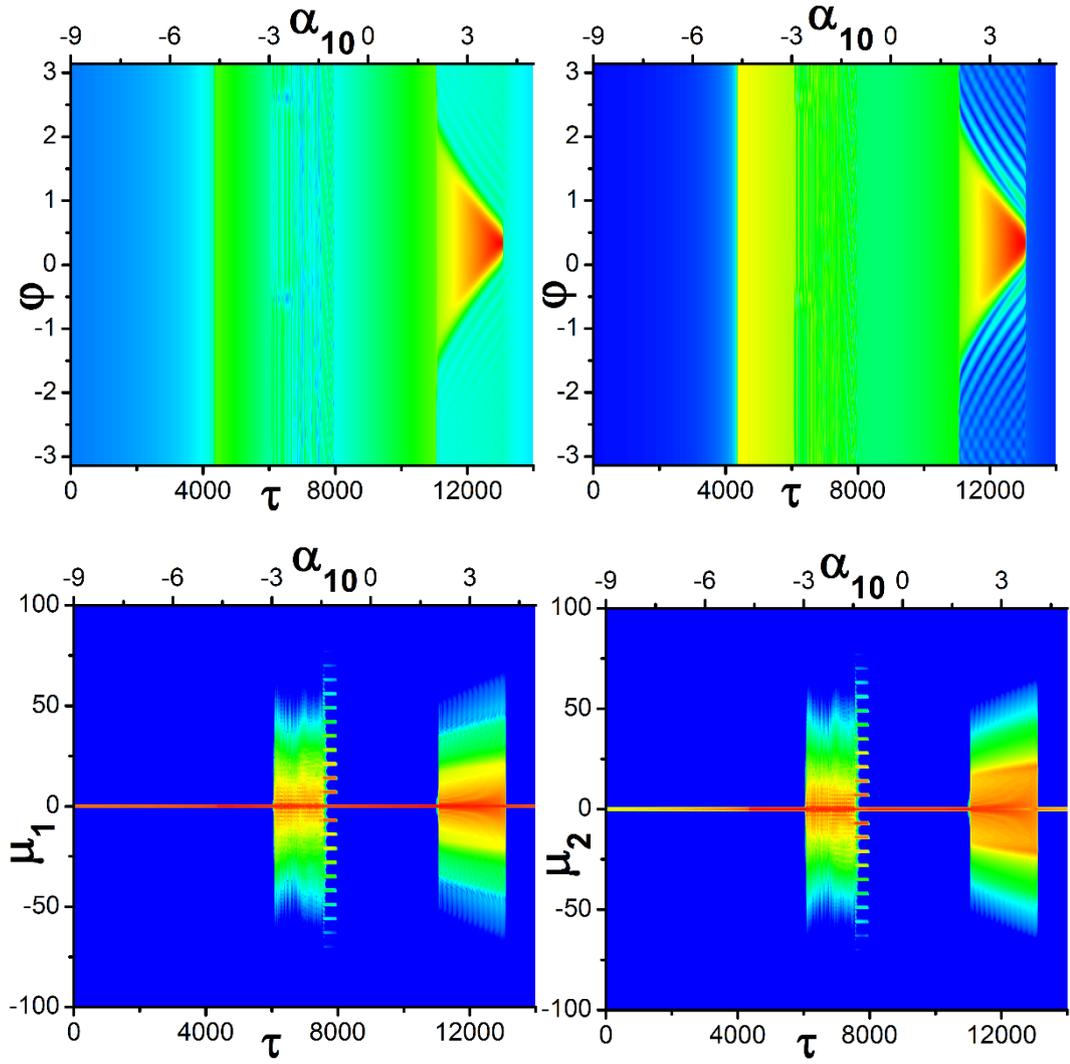

**Fig. 11.** (left column) FW and (right column) SH field distribution (upper line) and spectrum (bottom line) evolution upon forward frequency scan at $b_{21} = 0.05$, $b_{22} = -0.05$, $\Delta = \kappa_1$. All quantities are plotted in dimensionless units.

The dependence of the platicon generation domain position on the pumped mode shift is rather tricky and is not the same for positive and negative values of $\Delta$ (see Fig. 12, left panel). Nevertheless, it is clear, that shift absolute value must exceed some critical value. Besides, it should be noted, that while amplitude modulation provides the generation of the single platicon, in some cases mode coupling leads to the generation of several solitonic pulses. For example, in regions 1, 3, and 5 in the left panel in Fig. 12 we observed mostly the generation of the single platicon, while in regions 2 and 4 generation of several platicons (two or three) is more possible. Generation domain shifts to the larger absolute values of the detuning with the growth of the pump amplitude (see Fig. 12, right panel). Regions 1 and 3 corresponds to the predominant generation of the single platicon, while in region 2 multi-platicon generation is more frequent.

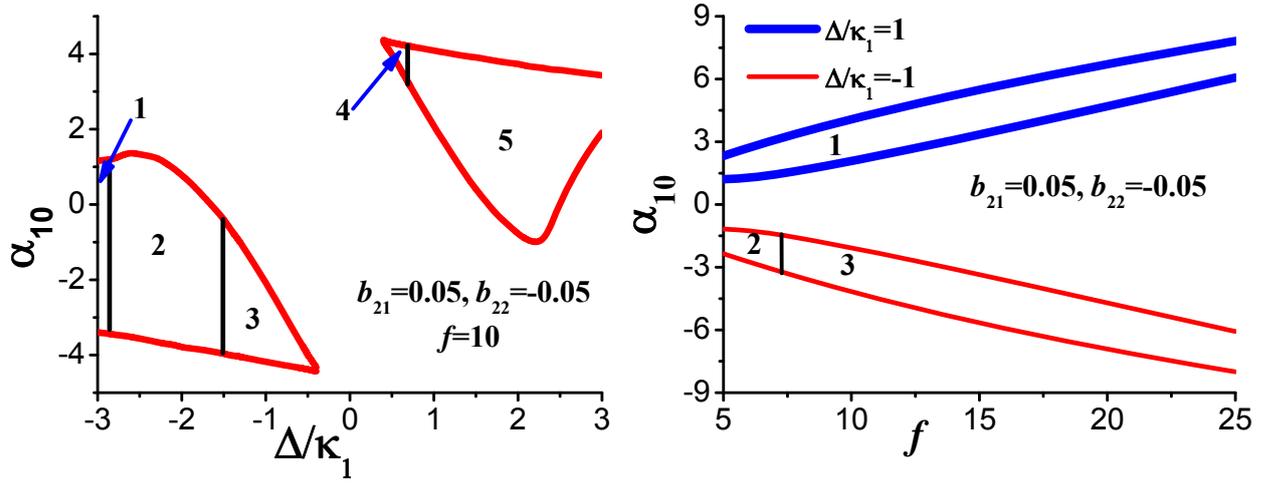

**Fig. 12.** Platicon generation domains for different values of the pumped mode shift and fixed pump amplitude (left panel) and different pump amplitudes and fixed pumped mode shifts (right panel) at $b_{21} = 0.05$, $b_{22} = -0.05$. Regions 1, 3, 5 in left panel and 1, 3 in right panel corresponds to the domains of the predominant single platicon generation. All quantities are plotted in dimensionless units.

It was found, that for the same absolute value of the effective pump detuning $\alpha_{10}$ amplitudes of platicon components at positive and negative $\alpha_{10}$ are almost equal, but at positive detunings SH component possesses more oscillating tails, while at negative detunings – FW component (see Fig. 13).

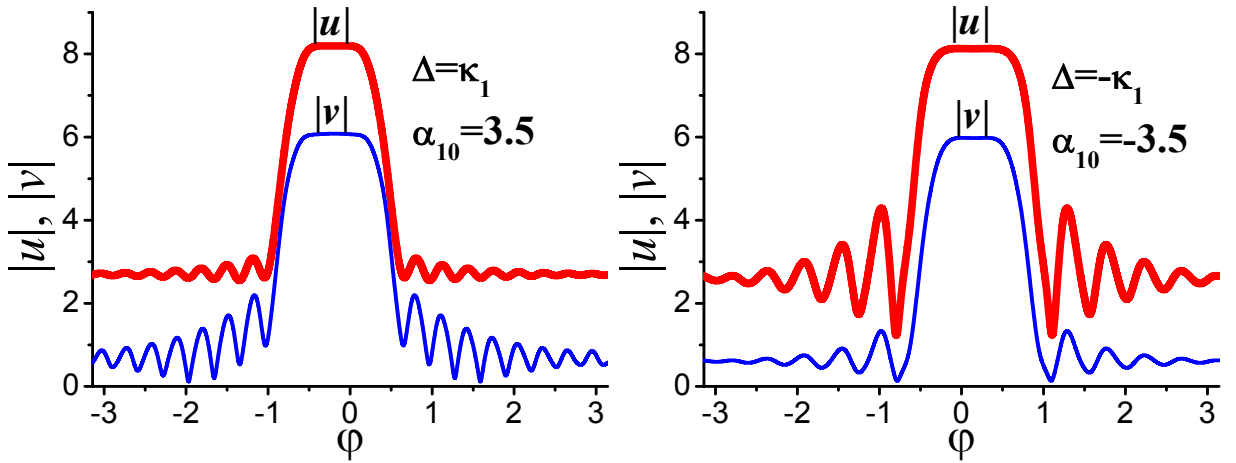

**Fig. 13.** Profiles of the platicon components for the different pumped mode shift values at $b_{21} = 0.05$, $b_{22} = -0.05$. All quantities are plotted in dimensionless units.

For the opposite GVD signs (at $b_{21} = -0.05$, $b_{22} = 0.05$) the situation is practically the same (see Fig. 14) except platicon tails: at positive detunings FW component possesses more oscillating tails, while at negative detunings – SH component (compare left panel in Fig. 13 and right panel in Fig. 14).

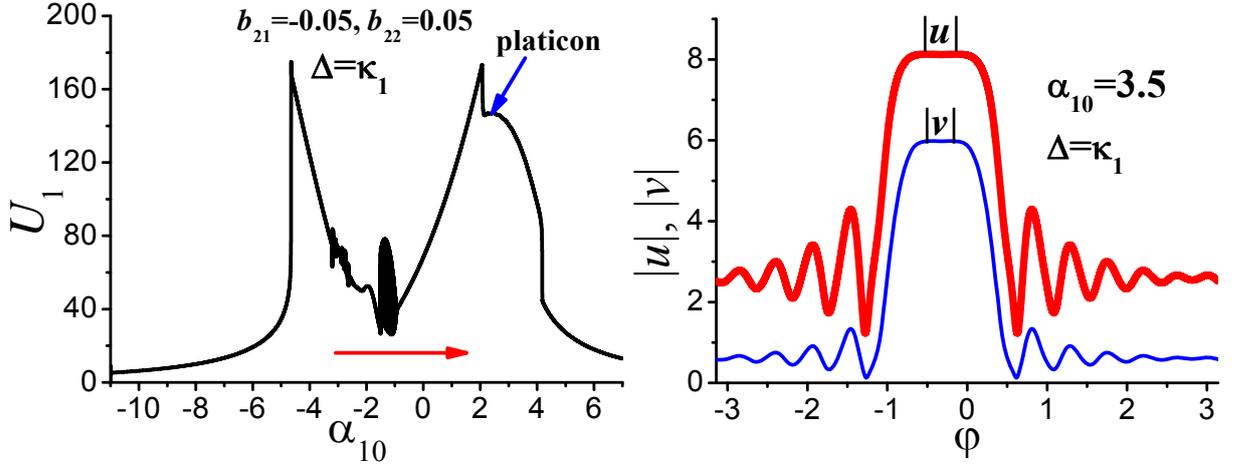

**Fig. 14.** (left panel) The dependence $U_1(\alpha_{10})$ upon forward scan and (right panel) profiles of the platicon components for $\alpha_{10} = 3.5$ at $b_{21} = -0.05$, $b_{22} = 0.05$, $\Delta = \kappa$. All quantities are plotted in dimensionless units.

As for the generation domains at opposite combinations of the GVD coefficients, it was found that domains also coincide when replacing simultaneously $\alpha_{10}$ by $-\alpha_{10}$ and $\Delta$ by $-\Delta$ (compare Fig. 12 and 15).

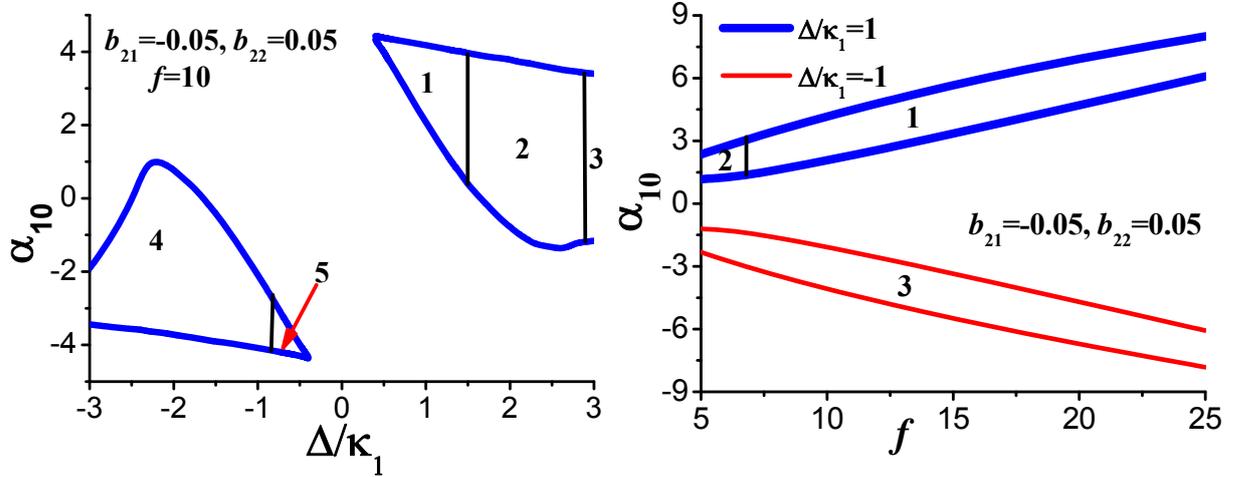

**Fig. 15.** Platicon generation domains for different values of the pumped mode shift and fixed pump amplitude (left panel) and for different pump amplitudes and fixed pumped mode shifts (right panel) at $b_{21} = -0.05$, $b_{22} = 0.05$. Regions 1, 3, 4 in left panel and 1, 3 in right panel corresponds to the domains of the predominant single platicon generation. All quantities are plotted in dimensionless units.

We also checked that this method is applicable for the wide range of the material and pump parameters, such as pump power and GVD coefficients. Similar to the previous method, absolute values of the GVD coefficients should be rather close (for $f=10$, $\Delta=\kappa_1$ platicon generation was observed if $0.7 \leq |b_{21}/b_{22}| \leq 3.0$ at $b_{21}=0.05$ and $0.4 \leq |b_{21}/b_{22}| \leq 2.0$ at $b_{21}=-0.05$). This approach is also sensitive to the matching of FSRs at the fundamental and at second harmonic frequencies and may be applied if FSRs difference is small enough (the critical value of the FSRs difference is about $\kappa_1$ for $\Delta=\kappa_1$). In that case the platicon repetition rate may differ from FSRs at the fundamental wave and the second harmonic due to the nonlinear effects [32, 38]. Also, there is a finite range of the resonant frequency offset $\delta$ providing platicon generation. It depends on the value of the pump mode shift and on the GVD coefficients and is asymmetric with respect to the point $\delta=0$. For example, at $f=10$, $\Delta=\kappa_1$, $d=0$ generation of the platicons is possible if $-10<\delta<2$ at $b_{21}=0.05$, $b_{22}=-0.05$ (see Fig. 16) and $-2<\delta<10$ at $b_{21}=-0.05$, $b_{21}=0.05$.

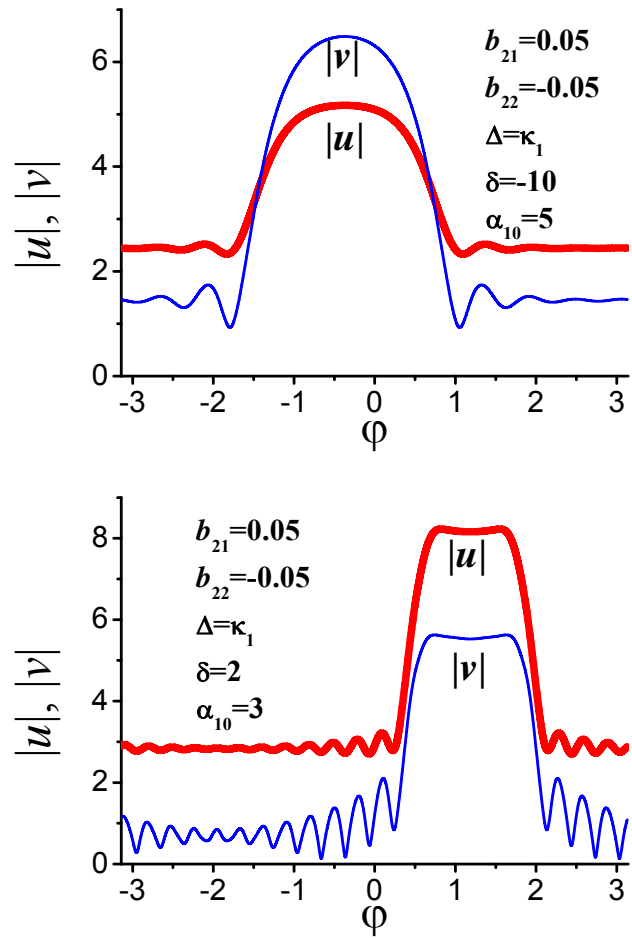

**Fig. 16.** Profiles of the platicon components at $f=10$, $d=0$, $\Delta=\kappa_1$, $b_{21}=0.05$, $b_{22}=-0.05$ for $\delta=-10$, $\alpha_{10}=5$ (left panel) and $\delta=2$, $\alpha_{10}=3$ (right panel). All quantities are plotted in dimensionless units.

In conclusion, we demonstrated for the first time that under particular conditions (e.g. if the difference between the FSRs at the fundamental and at second harmonic frequencies is small enough and the condition of resonant frequencies matching is almost satisfied) it is possible to generate two-color flat-top solitonic pulses, platicons, in $\chi^{(2)}$ microresonators using pump amplitude modulation or controllable mode interaction approach, if GVD signs at the fundamental frequency and its second harmonic are opposite. Platicon generation processes were simulated numerically, and generation conditions and excitation domains were found for the both approaches and both combinations of the GVD coefficients. Previously unknown generation regimes were discovered: unpredictably, platicon generation was observed on both sides of the linear resonance at both positive and negative pump detunings. Properties of the generated patterns were studied for the different combinations of the medium parameters. We expect that our results will be useful for the understanding the dynamics of the nonlinear processes in quadratic microresonators and will contribute to the efficient generation of the coherent frequency combs in such systems. $\chi^{(2)}$ microresonators provide unique opportunity of platicon generation with substantially decreased pump power compared to Kerr combs, in spectral regions where Kerr platicons are hardly possible and with high pump-to-comb conversion efficiency that is particularly promising for important up-to-date applications, for example, such as coherent optical communications.

This work was supported by the Russian Science Foundation (Project 17-12-01413).